\def\l{\lambda }  \def\r{\varrho }  \def\s{$\,$}
\font\tenyyy=cmcsc10 \def\yyy{\tenyyy}
\documentstyle[12pt]{article}
\newcommand{\beq}{\begin{equation}}
\newcommand{\beqar}{\begin{eqnarray}}
\newcommand{\eeq}[1]{\label{#1} \end{equation}}
\newcommand{\eeqar}[1]{\label{#1} \end{eqnarray}}
\begin{document}
\begin{titlepage}
\centerline{{\huge Chaos in Axially Symmetric Potentials with }}
\centerline{{\huge Octupole Deformation}}
\medskip
\centerline{{\yyy
W.D.\s Heiss$^{\star}$ , R.G.\s Nazmitdinov$^{\star \star}$
\footnote{on leave of absence from
Joint Institute for Nuclear Research,
Bogoliubov Laboratory of Theoretical Physics, 141980 Dubna, Russia}
and S.\s Radu$^{\star}$ }}
\medskip
\centerline{{\sl
$^{\star}$ Centre for Nonlinear Studies and Department of Physics}}
\centerline{{\sl
University of the Witwatersrand, PO Wits 2050, Johannesburg, South Africa }}
\centerline{{\sl
$^{\star \star}$ Departamento de Fisica Teorica C-XI }}
\centerline{{\sl Universidad Autonoma de Madrid, E-28049, Madrid, Spain}}
\baselineskip 20pt minus.1pt
\begin{abstract}
Classical and quantum mechanical results are reported for the single particle
motion in a harmonic oscillator potential which is characterized by a
quadrupole deformation and an additional octupole deformation. The chaotic
character of the motion is strongly dependent on the quadrupole deformation
in that for a prolate deformation virtually no chaos is discernible while
for the oblate case the motion shows strong chaos when the octupole term
is turned on.
\end{abstract}
\end{titlepage}
\newpage
\textheight=23cm
\voffset=-2.2cm
The mean field approach is of central importance in any theoretical
description of a many body system. In nuclear physics it is the basis for
the shell model and its extensions such as collective states. Its success
has been further demonstrated in the application to deformed nuclei and to
metallic clusters where
spherical symmetry is given up because of experimental evidence. Usually,
quadrupole deformation is considered to be the major deviation from spherical
symmetry. However, more recently a possible octupole contribution has been
taken into account for a number of reasons \cite{A90,Ha91}.

Inclusion of an octupole term in addition to a quadrupole term renders the
classical single particle motion nonintegrable. In fact, the system turns
out to be chaotic. This has been discussed by a number of authors with
various degrees of simplification \cite{Ari93}. A nicely systematic approach
is given in \cite{Ar87}, where the motion in a quadrupole deformed cavity is
analyzed and the terms that give rise to actual chaotic behaviour
are clearly distinguished. In a recent investigation \cite{Blocki93}
the study of classical motion in a cavity with oscillating walls of
even and odd higher
order multipoles has led to interesting conclusions about elastic versus
dissipative behaviour of a noninteracting gas depending on the integrability
or nonintegrability of the equations of motion. Surfaces of section are
used to discern the onset and degree of chaotic motion.

This paper is similar in spirit in that we investigate a simplified
model where we leave out terms, which, albeit physically important, are prone
to blur the analysis when the interest is focussed on the essentials that
give rise to chaotic behaviour. Since our aim is directed not only to the
classical but also to the corresponding quantum mechanical motion,
we leave out the spin-orbit term and the ${\stackrel{\rightarrow}{l}}^2$-term
pr
Nilsson model to render as closely as possible the analogy between the
classical and quantum case. The importance of the
${\stackrel{\rightarrow}{l}}^2$-term is well known in
nuclear physics and for metallic clusters\cite{Heer93}. To determine
its role in the context of chaotic motion the corresponding
classical case ought to be studied. We defer treatment of this term.
Despite the simplifications, such a model allows to
understand the main features of shell structure effects, for example, in
super--(hyper) deformed nuclei\cite{BM,Naz92}.
Recent experimental data of superdeformed $K$-isomers in nuclei \cite{Dele93}
and electronic shell structure effects in metallic clusters \cite{Heer93}
clearly underline the importance of oblate deformation. Therefore we
investigate in the present paper the effect of the octupole term  for
prolate and oblate deformation; the oblate/octupole case has not been dealt
with in ref.\cite{Blocki93}. We analyse the case of
zero temperature which is good for nuclei.
For metallic clusters finite temperature should be considered\cite{Martin91};
however, our interest is focussed on shell structure whose character is
unaffected by temperature except for the amplitude.
Only axially symmetrical terms are taken into account which
brings down to two the number of degrees of freedom of the motion.

The single particle motion is considered in the potential
\begin{equation}
V(\r  ,z) = \frac{m \omega ^2}{2}\biggl ( \r ^2+{z^2\over b^2}  +
\l  \frac{2z^3-3z \r  ^2}{\sqrt{\r  ^2+z^2}}\biggr )
\end{equation}
where $\r  ^2=x^2+y^2$ in cartesian coordinates $x,y,z$. We
recognize an axially symmetric harmonic oscillator with frequencies
$\omega_x = \omega_y = b \omega_z$
 with an octupole term, written in the cylindrical coordinates $(\r ,
z,\phi)$. For the results presented the parameters are chosen such that
the levels are 15MeV apart for $b=1$ and $\l  =0$ which corresponds to a
three dimensional isotropic harmonic oscillator. For $\l  =0$ and
$b>1\left(b<1\right)$ we have the mean field potential of a prolate (oblate)
nucleus. Note that, choosing a suitable set of different parameters, we deal
with a metallic cluster.  The octupole term ($\l  \ne 0$)
is proportional to $r^2 Y_{30}$ with $Y_{30}$ being the third order spherical
harmonic and $r^2=\r  ^2+z^2$.

If $|\l  |<\l  _{\rm c}$ we are dealing with a proper bound
state
problem. Here $\l  _{\rm c}$ is defined to be the value for which
the potential no longer binds, for $|\l |>\l _{\rm c}$ the potential
tends to $-\infty $ along one or two directions. The direction and the
value of $\l  _{\rm c}$ depend on the quadrupole deformation $b$.
For prolate nuclei ($b>1$), $\l  _{\rm c}=1/(2b^2)$ and the
potential opens its valley along the positive (negative) $z$-direction for
negative
(positive) $\l $. For oblate nuclei ($0.5\le b<1$) the other possible
direction which is along the line
$\r  ={\rm sign}(\l  )\alpha z$, with $\alpha \approx 0.4$,
is of increasing importance. At the value
$b\approx 0.58$ valleys along the two directions $\r  =0$ and
$\r  =0.4z$
open simultaneously for $\l  _{\rm c}\approx 1.5$ while for a still
smaller
value of $b$, say for $b=0.5$, the valley along the direction $\r  =0.4z$
opens for $\l  _{\rm c}\approx 1.64$ while the one along the $z$-axis
now opens for a larger value of $\l  $. (Analytic expressions for
$\l  _{\rm c}$ and the direction $\alpha $ as functions of $b$ exist
but are of little interest.)
Chaotic motion is expected to become the more pronounced the nearer the
parameters for a bounded motion are to those for an
unbounded motion. Thus increasing chaotic behaviour
is expected when $\l  $ approaches $\l  _{\rm c}$. Furthermore,
 since
also the geometry of the potential
for $\l  =\l  _{\rm c}(b)$ depends on $b$, the chance of an escape
 for
$\l  \ge \l  _{\rm c}$ also depends on $b$. This chance reflects
upon the amount of chaos, i.e.\s the Lyapunov exponent, for the bounded motion
prevailing for values $\l  <\l  _{\rm c}(b)$. In this way the
amount of chaos is expected to
be greatest for $b\approx 0.58$ where two valleys can serve as an escape
route if $\l  =\l  _{\rm c}$.
In addition, at $b=2$ (superdeformed prolate nucleus) less chaos is
expected than
at $b=0.5$ (superdeformed oblate nucleus), since the opening of the
phase space happens
through a small bottle neck in the former case thus considerably reducing
the chance for escape, while the chance for escape at $b=0.5$
is enhanced as the phase space opens more widely.

The results of the numerical integration of the equations of motion
confirm all the expectations. Axial symmetry yields the constant of motion
$p_{\phi }=\r  ^2 \dot \phi $ of the three dimensional motion.
We present results only for $p_{\phi } =0$ . A non-zero value of the
$z$-component of the angular momentum does not produce new insights.
Since the potential scales as $V(\gamma \vec r)=\gamma ^2 V(\vec r)$ it
suffices to investigate one energy only \cite{Boh}.

For the superdeformed prolate nucleus ($b=2$) there is hardly any chaotic
behaviour discernible in the classical motion for all $\l  <\l  _{{\rm
 c}}$.
When looking at surfaces of section which we have taken at $\r  =0$
 (recall
$p_{\phi }=0$) in the $(p_z-z)$-plane there is some scattering of the dots
when $\l  $ is very close to $\l  _{\rm c}$, but the Lyapunov
 exponent
is virtually zero. There is, however, a proliferation of periodic orbits, an
aspect important below when we discuss quantum mechanics. This is illustrated
in Fig.(1a) where surfaces of section are displayed for $b=2$ and
$\l=2/3 \l _{{\rm c}}$.

For decreasing values of $b$ the motion becomes increasingly chaotic up to the
maximally chaotic case at $b\approx 0.58$. Surfaces of section are illustrated
in Fig.(1b) for $\l=1/3 \l _{{\rm c}}$. Note, that for $b=0.58$, no
structure would be discernible for $\l=2/3 \l _{{\rm c}}$. We
have compared results for $\l  =\beta \l  _{\rm c}(b)$ with $\beta
 = $
0.2, 0.4, 0.6 and 0.9. The trend is uniform
in that the Lyapunov exponent shows the behaviour as indicated in Fig.(2).
The figure presents values referring to chaotic orbits and avoids initial
conditions within regions of stability. Such regions of stability gradually
disappear when $\l  $ approaches $\l  _{\rm c}$ if $b\le 1$.

We find the expected proliferation of periodic orbits. However, we refrain
from discussing them in great detail as the aspect of the possible retrieval
\cite{Win89} of periodic orbits from the quantum mechanical spectrum is
discussed in a forthcoming paper.

The quantum mechanical treatment is straightforward in principle.
In the spirit
of previous work \cite{He} we use for the full problem which is of the form
$H_0+\l  H_1$ a representation where $H_0$ is diagonal.
The basis chosen is referred \cite{BM} to as the basis
using the asymptotic quantum numbers $n_{\perp },n_z$ and $\Lambda $ where
$n_{\perp }=n_++n_-$. For a fixed value of $\Lambda $
this leaves two quantum numbers (reflecting the two degrees of freedom)
to enumerate the rows and columns of the matrix problem. For $\Lambda =0$ the
diagonal entries of $H_0$ are thus $E_{n_{\perp },n_z}=
\hbar \omega (n_{\perp } +1+(n_z+1/2)/b)$. The matrix elements
of $H_1$ are obtained from those of $z \sim (a^{\dagger }_z+a_z)$ and
$\r  ^2\sim (A_+A_+^{\dagger } +A_-A_-^{\dagger }+A_+A_-+A_+^{\dagger }
A_-^{\dagger })$, where, in terms of the usual boson operators $a_{x,y}$,
$A_{\pm}=(a_x\mp i a_y)/\sqrt{2}$.
The effect of truncation was tested by looking at
the variation of the lower end of the spectrum when the dimension of the
matrices was increased. There is certainly a dependence on $b$ and $\l  $.
 For
$\l  \le 0.9\l  _{\rm c}$ and $0.5\le b \le 2$ the variation was
 less than
$1 \%$ for the first 300 levels obtained from 1600$\times $1600 dimensional
matrices.

For demonstration we display parts of spectra in Figs.(3). It is obvious,
and in fact quantitatively confirmed in Figs.(4) where the relevant
statistical analyses are shown, that the quantum mechanical results are in
line with the classical cases. The level repulsions in the superdeformed
prolate case are very weak thus giving rise to a nearest neighbor
distribution (NND) which appears nearer to an integrable case than to the
typical Wigner distribution. Of physical interest are the pronounced new shell
structures that emerge near to $\l  =0.5 \l  _{\rm c}$
and $\l  =0.65 \l  _{\rm c}$ in the superdeformed prolate case. This
pattern is directly related to the periodic orbits indicated in Fig.(1a)
as will be discussed in detail in a forthcoming paper.
Since we have left out terms like spin-orbit
coupling and ${\stackrel{\rightarrow}{l}}^2$ we cannot claim that such
structure
where we find them. However, the essential point is the fact that such
structures will always emerge. Since the situation is so close to
integrability for all $\l  < \l  _{\rm c}$, the level
repulsions are always weak; as a consequence, shell structures are bound to
emerge for some values of the parameters. We note, that the consequent
periodicity in the spectrum will produce sharp lines in the Fourier transform
of the level density \cite{hemu}, a reflection of the many periodic orbits
\cite{Gutz} found in the corresponding classical case as was mentioned above.

These findings nicely contrast with the oblate case where the spectrum and
in particular the NND have all signatures of chaos. For sufficiently large
values of the octupole strength $\l  $ all periodic structure is destroyed
and there is no scope for new magic numbers. We mention that the NND best
approaches the Wigner surmise for $b\approx 0.58$.

In summary, we have found for a problem typical for nuclear physics and
for metallic clusters that chaotic behaviour should be
expected in principle even for the single particle motion if
deformations of higher order than quadrupole are taken into account. Moreover,
for the prolate and in particular the superdeformed prolate case, there is a
remarkable stability against chaos when octupole deformation is switched on.
This result is in agreement with a prediction of an octupole instability
of super-- and hyperdeformed nuclei \cite{Ab90}, based on realistic
calculations. We conjecture that this
pattern prevails also when axial symmetry is broken,
i.e.\s when terms of the form $Y_{3m}$, $m \ne 0$ are taken into account. To
what extent stability against chaos can be associated with stability in
general terms, - it is a well known fact that there are more prolate than
oblate nuclei -, is subject to further investigation. In the spirit of
ref.\cite{Blocki93} we should expect prolate nuclei to behave rather
elastically in contrast to oblate nuclei where more dissipative behaviour
is anticipated. We may speculate that the absence of
shell structure for the oblate/octupole case could
prevent the existence of stable oblate/octupole deformed clusters.

\vskip 1cm

{\bf Acknowledgments.} Two of the authors (WDH and RGN) met at the workshop
on Microscopic Nuclear Structure at the INP, Seattle during September/October
1992 where the conducive working atmosphere led to this collaboration. RGN
acknowledges financial support from DGICYT of Spain. We are
also grateful for discussions with Dr.\s David Sherwell on aspects
of classical chaos.

\newpage

\centerline{Figure Captions}

\vspace{0.5cm}

{\bf Fig.1} Surfaces of section at $\r =0$
for $b=2,\,\l =2/3\l _{{\rm c}}$ (left) and
$b=0.58,\,\l =1/3\l _{{\rm c}}$ (right). In the left part stability islands
are clearly discernible for winding numbers 2:5, 1:2 and 4:7. The right part
is dominated by chaotic motion; some of the remaining islands are indicated
by ellipses.

\vspace{0.5cm}

{\bf Fig.2} Lyapunov exponent as a function of $\l /\l _{\rm c}$. The
lower solid line is for $b=1$, the upper solid line for $b=0.58$ and the
dashed line for $b=0.5$. The values for $b=2$ are too small to appear on the
diagram.

\vspace{0.5cm}

{\bf Fig.3} A section of energy levels as a function of
$\l /\l _{\rm c}$ for $b=2$ (top) and $b=0.5$ (bottom).

\vspace{0.5cm}

{\bf Fig.4} Nearest neighbor distribution of the spectra shown in Fig.2
for $\l /\l _{\rm c}=0.5$ as a function of the unfolded energy.
For the top ($b=2$) 200 levels and for the bottom ($b=0.5$)
600 levels have been used.

\end{document}